\documentclass{elsart}

\usepackage{graphicx}

\usepackage{amssymb}

\begin{document}

\begin{frontmatter}



\title{Effects of mixed rare earth occupancy on the low temperature properties of (R, R',R''...)Ni$_{2}$Ge$_{2}$ single crystals}


\author{S. A. Law, S. L. Bud'ko, P. C. Canfield}

\address{Ames Laboratory and Department of Physics and Astronomy, Iowa State University, Ames, IA 50011, USA}

\begin{abstract}
Temperature and applied magnetic field dependent magnetization measurements on 34 single crystalline samples of
(R, R',R''...)Ni$_{2}$Ge$_{2}$ compounds (R, R', R'', etc. being primarily Gd$-$Lu, Y), were made. These measurements
reveal that, despite extremes in local moment anisotropy, the average de Gennes parameter is a remarkably good
predictor of the paramagnetic to antiferromagnetic ordering temperature.  In addition, the pronounced metamagnetic
phase transitions seen in the low temperature phase of TbNi$_{2}$Ge$_{2}$ are found to be remarkably robust to
high substitution levels of Gd and 25\% substitutions of other heavy rare earths.
\end{abstract}

\end{frontmatter}

\section{Introduction}

The RNi$_{2}$Ge$_{2}$ series is one of the model series for the study of how the 4f-shell electrons manifest
themselves in intermetallic compounds.  Single crystals can be grown for all stable R isotopes (R = La$-$Nd,
Sm$-$Lu, and Y) and very likely could be made for R = Pm as well.\cite{1}  This range of R allows for the
formation of non-magnetic (R = La, Lu, and Y), hybridizing (R = Ce and Yb), isotropic local moment (R = Eu and
Gd), extremely axial local moment (R = Tb), and extremely planar local moment systems (R = Er and Tm) as well as a
range of less extreme, but still anisotropic examples.  All of the local moment members of the RNi$_{2}$Ge$_{2}$
series manifest antiferromagnetic order below \textit{T$_{N}$}, and the \textit{T$_{N}$} values for the heavy rare
earths scale well with the de Gennes parameter (dG = (g$_{J}$-1)$^{2}$J(J+1)).  In addition to an extensive study of
the anisotropic thermodynamic and transport properties of the pure RNi$_{2}$Ge$_{2}$ series,\cite{1} studies of
the magnetic structures in GdNi$_{2}$Ge$_{2}$ and TbNi$_{2}$Ge$_{2}$,\cite{2,3,4} the evolution of an Ising spin
glass state in (Y$_{1-x}$Tb$_{x}$)Ni$_{2}$Ge$_{2}$,\cite{5} and the effects of band filling on the ordering wave
vector in EuNi$_{2}$Ge$_{2}$ and GdNi$_{2}$Ge$_{2}$ have been made.\cite{6,7}

Rare earth intermetallic compounds offer the possibility of utilizing the intelligible complexity of the 4f-shell to tune magnetic properties.  By changing the rare earth element, the size and anisotropy of the local moment as well as the value of the transition temperature can be changed.   The ability to modify the low temperature properties of this series by partial substitution of two or more R species onto the one unique crystallographic site offers the possibility of either tuning the magnetic properties with greater control or perhaps even creating new properties.  In this paper, we present the results of a broad study of the properties of (R, R',R''...)Ni$_{2}$Ge$_{2}$ single crystals, specifically focused on their low temperature local moment magnetism.

\section{Experimental methods}

Single crystals of (R, R',R''...)Ni$_{2}$Ge$_{2}$ were grown out of a self-flux rich in Ni and Ge.\cite{1,5}
Generally the ratio of starting elements was R$_{0.07}$Ni$_{0.465}$Ge$_{0.465}$ with the 7\% atomic for the rare
earth split appropriately between R, R', R'', etc.  (e.g., the (Tb$_{0.75}$Y$_{0.25}$)Ni$_{2}$Ge$_{2}$ growth
contained Tb$_{0.525}$Y$_{0.175}$Ni$_{0.465}$Ge$_{0.465}$).  The high purity elements were placed into
Al$_{2}$O$_{3}$ crucibles and sealed in quartz ampoules under a partial atmosphere of Ar gas.  The quartz ampoules
were then put into box furnaces, heated to just under 1200$^\circ$C, and slowly cooled (generally over 75$-$100
hours) to 1000$^\circ$C, at which point they were removed from the furnace and the excess liquid was
decanted.\cite{8},\cite{9}  The single crystals had a plate$-$like morphology and could have dimensions as large
as 1 x 1 x 0.2 cm$^{3}$, although more typical dimensions were 0.3 x 0.3 x 0.1 cm$^{3}$.  The crystallographic
\textit{c}-axis was found to be perpendicular to the plates and well-defined facets allowed for the easy
identification of the in-plane orientation as well.

Temperature and field dependent magnetization measurements were taken on Quantum Design MPMS units for \textit{T}
$\geq$ 1.8 K and \textit{H} $\leq$ 7 T, and resistivity and specific heat measurements taken on the m\'elange
sample (see below) were performed on Quantum Design PPMS units.

The antiferromagnetic ordering temperature, \textit{T$_{N}$}, was determined from low field, temperature dependent
magnetization data by invoking the similarity in temperature dependence between \textit{C$_{p}$(T)} and
${d(MT/H)}/{dT}$\cite{10} near \textit{T$_{N}$}.  Features in \textit{M(H)} curves were associated with critical
magnetic fields by determining \textit{H$_{c}$} to be the field at which ${dM}/{dH}$ was maximal.  A given
sample's de Gennes value was the weighted average of the constituent rare earth ions' elemental de Gennes values.

\section{Experimental results}

\subsection{(Pr$_{0.11}$Nd$_{0.11}$Sm$_{0.11}$Gd$_{0.11}$Tb$_{0.11}$Dy$_{0.11}$Ho$_{0.11}$Er$_{0.11}$Tm$_{0.11}$)Ni$_{2}$Ge$_{2}$ m\'elange sample}

\begin{figure}[htbp]
  \centering
  \begin{minipage}[b]{6.8 cm}
    \includegraphics[width=6.8cm]{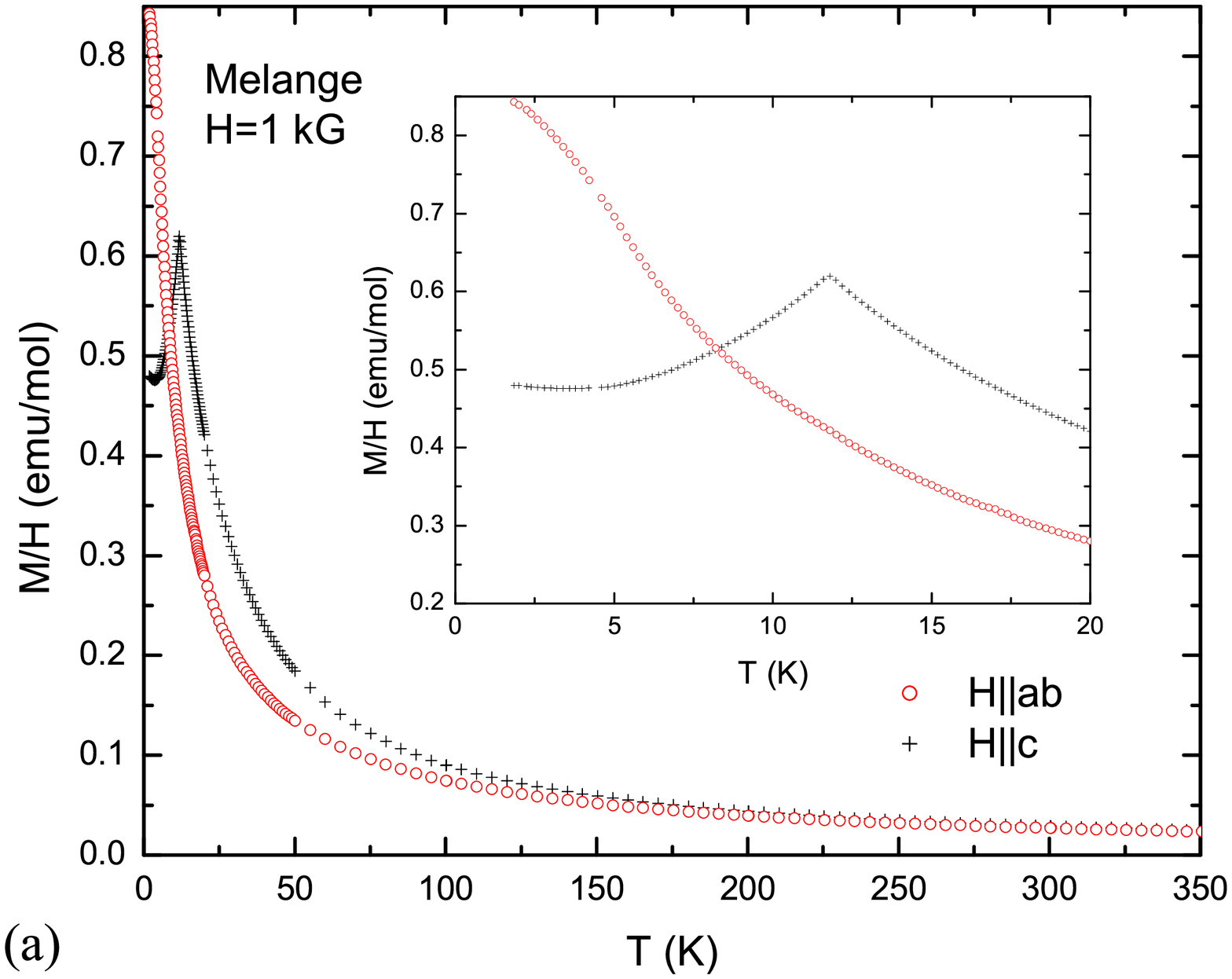}
  \end{minipage}
  \begin{minipage}[b]{6.8 cm}
    \includegraphics[width=6.8cm]{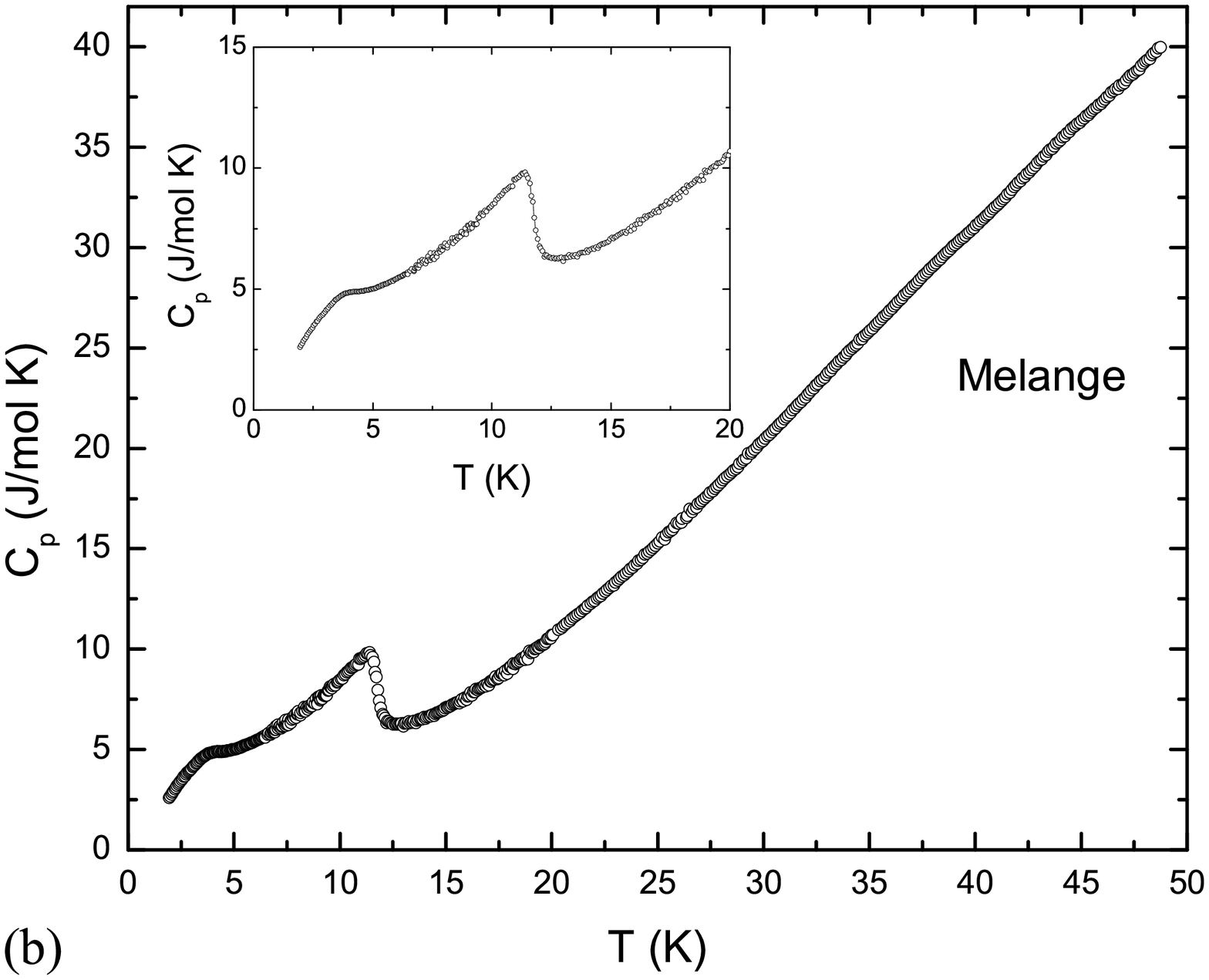}
  \end{minipage}
  \begin{minipage}[b]{6.8 cm}
    \includegraphics[width=6.8cm]{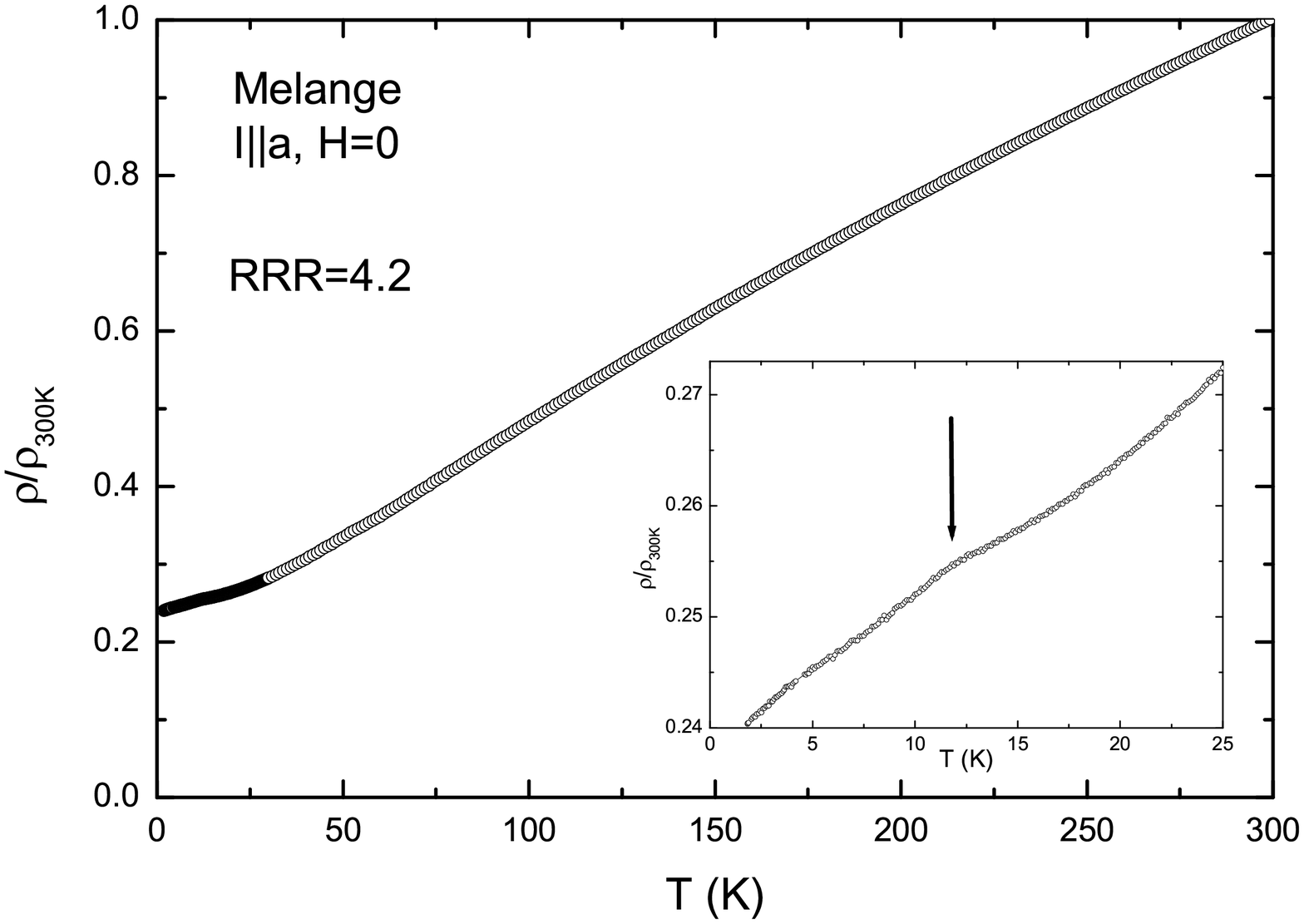}
  \end{minipage}
  \caption{Temperature dependent magnetic susceptibility (a), specific heat (b), and electrical reistivity (c) of the m\'elange sample, (Pr$_{0.11}$Nd$_{0.11}$Sm$_{0.11}$Gd$_{0.11}$Tb$_{0.11}$Dy$_{0.11}$Ho$_{0.11}$Er$_{0.11}$Tm$_{0.11}$)Ni$_{2}$Ge$_{2}$. The arrow in the inset to panel (c) marks the magnetic transition as observed in magnetization and specific heat.}
\end{figure}

\begin{figure}[htbp]
    \centering
    \includegraphics[width=6.8cm]{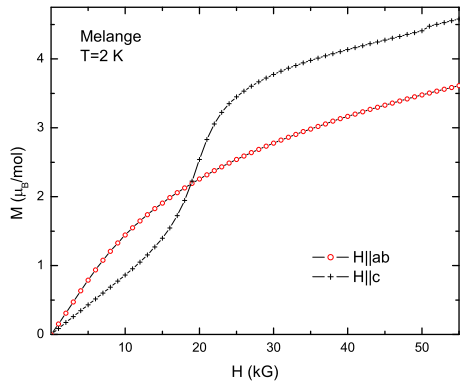}
    \caption{Anisotropic, low temperature, field dependent magnetization of the m\'elange sample, (Pr$_{0.11}$Nd$_{0.11}$Sm$_{0.11}$Gd$_{0.11}$Tb$_{0.11}$Dy$_{0.11}$Ho$_{0.11}$Er$_{0.11}$Tm$_{0.11}$)Ni$_{2}$Ge$_{2}$.}
\end{figure}

Single crystals of
(Pr$_{0.11}$Nd$_{0.11}$Sm$_{0.11}$Gd$_{0.11}$Tb$_{0.11}$Dy$_{0.11}$Ho$_{0.11}$Er$_{0.11}$Tm$_{0.11}$)Ni$_{2}$Ge$_{2}$
(henceforth referred to as the m\'elange sample) were grown to see if a thorough mixture of the single site local
moment anisotropy would result in a partial suppression of long-range order, or perhaps even a complete loss of
long range order resulting in a spin glass state akin to that associated with random site anisotropy.  Figure 1
presents the magnetic susceptibility, specific heat data, and electrical resistivity for this sample. Figure 2
presents the anisotropic magnetization as a function of applied field for \textit{T} = 2 K for \textit{H $||$ c}
plane and \textit{H $||$ ab} plane.  Both temperature dependent thermodynamic measurements reveal a phase
transition at 11.4 $\pm$ 0.5 K.  The transport data also reveal a somewhat weaker feature near this temperature.
This phase transition is consistent with long range antiferromagnetic order, and magnetic X-ray diffraction
measurements reveal that there is an ordering wave vector (0, 0, 0.777) in the low temperature state. \cite{11}

The m\'elange sample's transition temperature falls onto the de Gennes scaling line associated with the heavy rare earth members of the RNi$_{2}$Ge$_{2}$ series [see ref. 1 and figure 9 below] indicating that there is little or no suppression of magnetic ordering in the sample even though there is a wide range of local moment anisotropy.  This opens up the possibility of selectively mixing rare earths on this one unique crystallographic site to tune \textit{T$_{N}$} systematically.  In the rest of this paper, we will mix a variety of heavy rare earths to test how well de Gennes scaling holds and to also investigate the effects of this mixing on the details of \textit{M(T)} and \textit{M(H)} data.

\subsection{Mixed heavy rare earth samples}

    Table 1 presents a summary of the magnetic data on the 34 mixed rare earth samples studied in this work.  These samples can be divided and discussed in a number of ways.  The first sub-group is a (Tb$_{1-x}$Gd$_{x}$)Ni$_{2}$Ge$_{2}$ series with samples ranging from the Ising-like TbNi$_{2}$Ge$_{2}$ to the Heisenberg-like GdNi$_{2}$Ge$_{2}$.  Figure 3 presents anisotropic \textit{M(T)/H} data for representative Gd substitutions: x = 0.15 and x = 0.90; the axial nature of Tb is clearly observed across the whole range of finite \textit{x} values.  Figure 4 presents a \textit{T} vs. \textit{x} diagram for the low field ordering of the (Tb$_{1-x}$Gd$_{x}$)Ni$_{2}$Ge$_{2}$ series.  The highest points which plot the paramagnetic to antiferromagnetic transition temperature vary with \textit{x} in a roughly linear manner.  The next highest transition varies in a non-monotonic fashion, reaching a minimum near \textit{x} $\sim$ 0.33.  The third, lowest temperature phase transition was tentatively identified for pure Gd, \cite{1} but is much more clearly seen and separated for light Tb doping, finally disappearing totally as pure TbNi$_{2}$Ge$_{2}$ is approached.

\begin{table}

\begin{tabular}{l|c|c|c|c|c|c}

R-site composition &de Gennes factor   &T$_{m}$ (K)    &$\theta_{c}$(K) &$\theta_{ab}$(K) &$\theta_{ave}$(K)   &p$_{eff}$($\mu_{B}$)\\
\hline
Tb$_{0.85}$Gd$_{0.15}$ &    11.29   &18.3, 7.3  &8.4    &-39.5  &-18.2  &9.3\\
Tb$_{0.75}$Gd$_{0.25}$  &11.81  &18.8, 7.3, 4.1 &12.4   &-36.3  &-13.8  &9.3\\
Tb$_{0.66}$Gd$_{0.33}$  &12.13  &20.3, 6.1  &8.6    &-39.5  &-14.8  &9.3\\
Tb$_{0.5}$Gd$_{0.5}$    &13.13  &21.1, 7.6, 3.6 &2.1    &-31.7  &-17.8  &9.2\\
Tb$_{0.33}$Gd$_{0.66}$  &13.86  &22.1, 10.1, 4.3    &-13    &-30    &-21.2  &8.8\\
Tb$_{0.15}$Gd$_{0.85}$  &14.96  &24, 16 &-13    &-21.1  &-18    &8.6\\
Tb$_{0.1}$Gd$_{0.9}$    &15.23  &25.4, 17.6, 8.3    &-12.1  &-19.7  &-16.9  &8.3\\
Tb$_{0.05}$Gd$_{0.95}$  &15.49  &25.4, 17.1, 3.1    &-18.1  &-18.1  &-18.1  &8.3\\

Y$_{0.25}$Tb$_{0.75}$   &7.88   &11.3, 4.3 &20.3   &-41.4  &-11.6  &8.8\\
Gd$_{0.25}$Tb$_{0.75}$  &11.81  &18.8, 7.3, 4.1 1&2.4   &-36.3  &-13.3  &9.3\\
Dy$_{0.25}$Tb$_{0.75}$  &9.65   &13.8, 7.8  &11.5   &-35.3  &-14.5  &10.1\\
Ho$_{0.25}$Tb$_{0.75}$  &9.01   &11.8, 4.9  &10.8   &-29.1  &-12    &10.3\\
Er$_{0.25}$Tb$_{0.75}$  &8.52   &12.3   &8.8    &-39    &-18.4  &10\\
Tm$_{0.25}$Tb$_{0.75}$  &8.17   &12.6   &13.2   &-41.5  &-12.8  &9.6\\
Yb$_{0.25}$Tb$_{0.75}$  &7.96   &10.8, 5.3  &16.2   &-40.9  &-13.7  &9.3\\
Lu$_{0.25}$Tb$_{0.75}$  &7.88   &12.3, 7.4, 3.3 &13.1   &-43.5  &-13.3  &9.4\\

Gd$_{0.75}$Er$_{0.25}$  &12.45  &21.8, 7.1  &-16.5  &-9.3   &-9.3   &8.4\\
Gd$_{0.5}$Er$_{0.5}$    &9.15   &17.3, 5.6  &-13.3  &-7 &-10.5  &8.8\\
Gd$_{0.25}$Er$_{0.75}$  &5.85   &8.8, 4.6   &-21.6  &0.05   &-5.2   &9.3\\

Gd$_{0.5}$Tb$_{0.5}$    &13.13  &21.1, 7.6, 3.6 &2.1    &-31.7  &-17.8  &9.2\\
Gd$_{0.5}$Er$_{0.5}$    &9.15   &17.3, 5.6  &-13.3  &-7 &-10.5  &8.8\\
Gd$_{0.5}$Tm$_{0.5}$    &8.46   &15.8, 3.3  &-22.6  &-3.4   &-8.2   &8.2\\
Tb$_{0.5}$Er$_{0.5}$    &6.53   &7.5    &6.8    &-19.8  &-8.8   &9.6\\
Tb$_{0.5}$Tm$_{0.5}$    &5.84   &6.3    &5.6    &-22.7  &-10.4  &9\\
Dy$_{0.5}$Ho$_{0.5}$    &5.79   &4.8    &6.4    &-15.5  &-6.5   &10.7\\
Dy$_{0.5}$Tm$_{0.5}$    &4.13   &3.6    &0.78   &-10.7  &-5.8   &9.6\\
Ho$_{0.5}$Er$_{0.5}$    &3.53   &3.1    &-4.8   &-0.94  &-2.6   &10.2\\
Er$_{0.5}$Tm$_{0.5}$    &1.86   &2.1    &-30.1  &6.8    &-1.8   &8.8\\

Tb$_{0.9}$Lu$_{0.1}$    &9.45   &14.8, 8.8  &14.1   &-45.7  &-14.1  &9.8\\

Gd$_{0.33}$Tb$_{0.33}$Er$_{0.33}$   &9.50   &14.5   &-0.35  &-15    &-8.9   &9.2\\
Gd$_{0.33}$Tb$_{0.33}$Tm$_{0.33}$   &9.05   &15.7   &1.7    &-21.4  &-11.9  &8.9\\
Gd$_{0.33}$Er$_{0.33}$Tm$_{0.33}$   &6.43   &11.75, 4.0 &-20    &0.93   &-5.4   &8\\
Gd$_{0.25}$Tb$_{0.25}$Dy$_{0.25}$Ho$_{0.25}$    &9.46   &13.2, 2.2  &2.8    &-19.2  &-10.5  &10\\
M\'elange   &5.35   &11.8 &3.8  &-15.9  &-6.18 &8.2\\

\end{tabular}
\\
\caption{Summary of magnetization measurements on (R,R',R'',...)Ni$_{2}$Ge$_{2}$ single crystals}

\end{table}

\begin{figure}[htbp]
  \centering
  \begin{minipage}[b]{6.8 cm}
    \includegraphics[width=6.8cm]{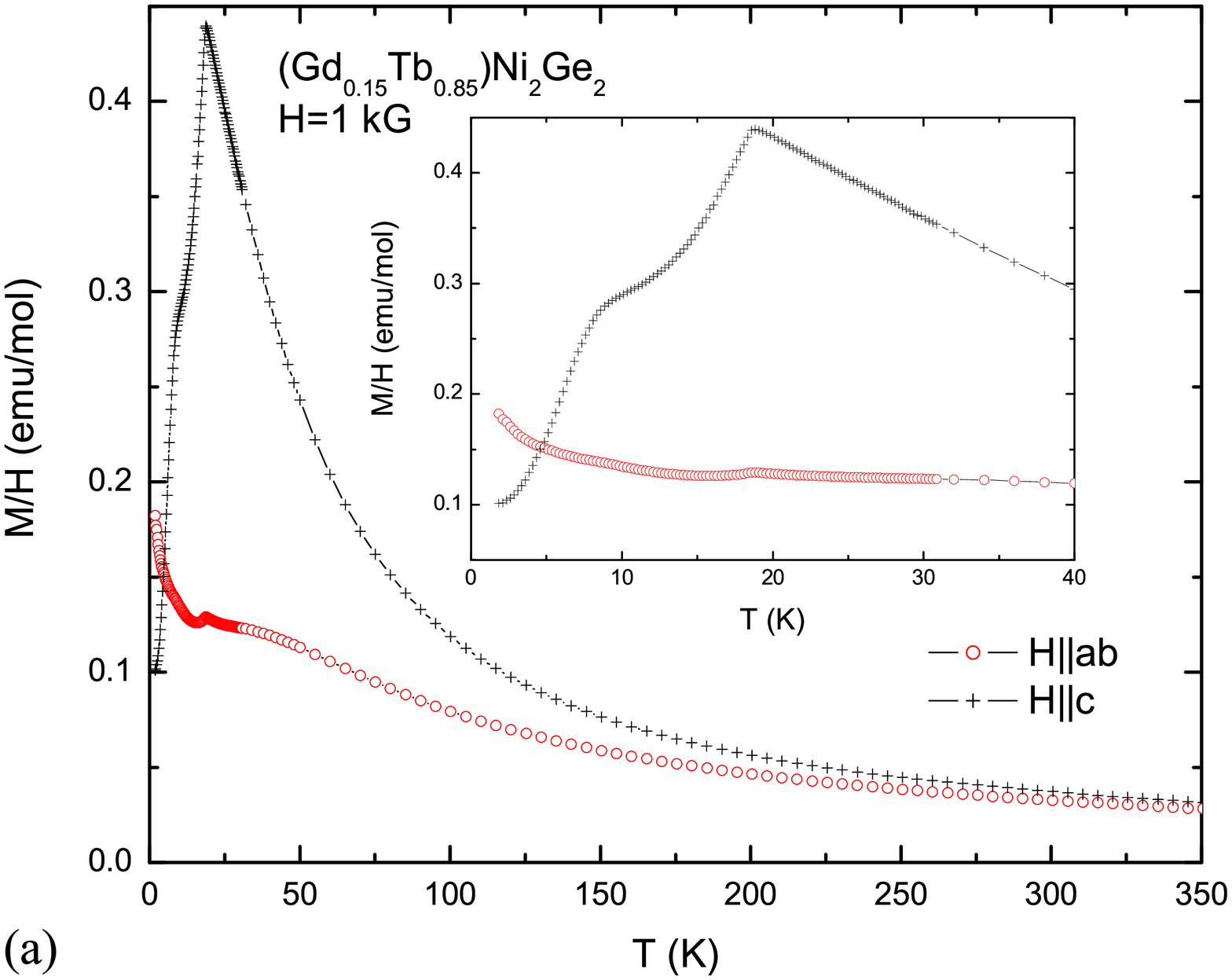}
  \end{minipage}
  \begin{minipage}[b]{6.8 cm}
    \includegraphics[width=6.8cm]{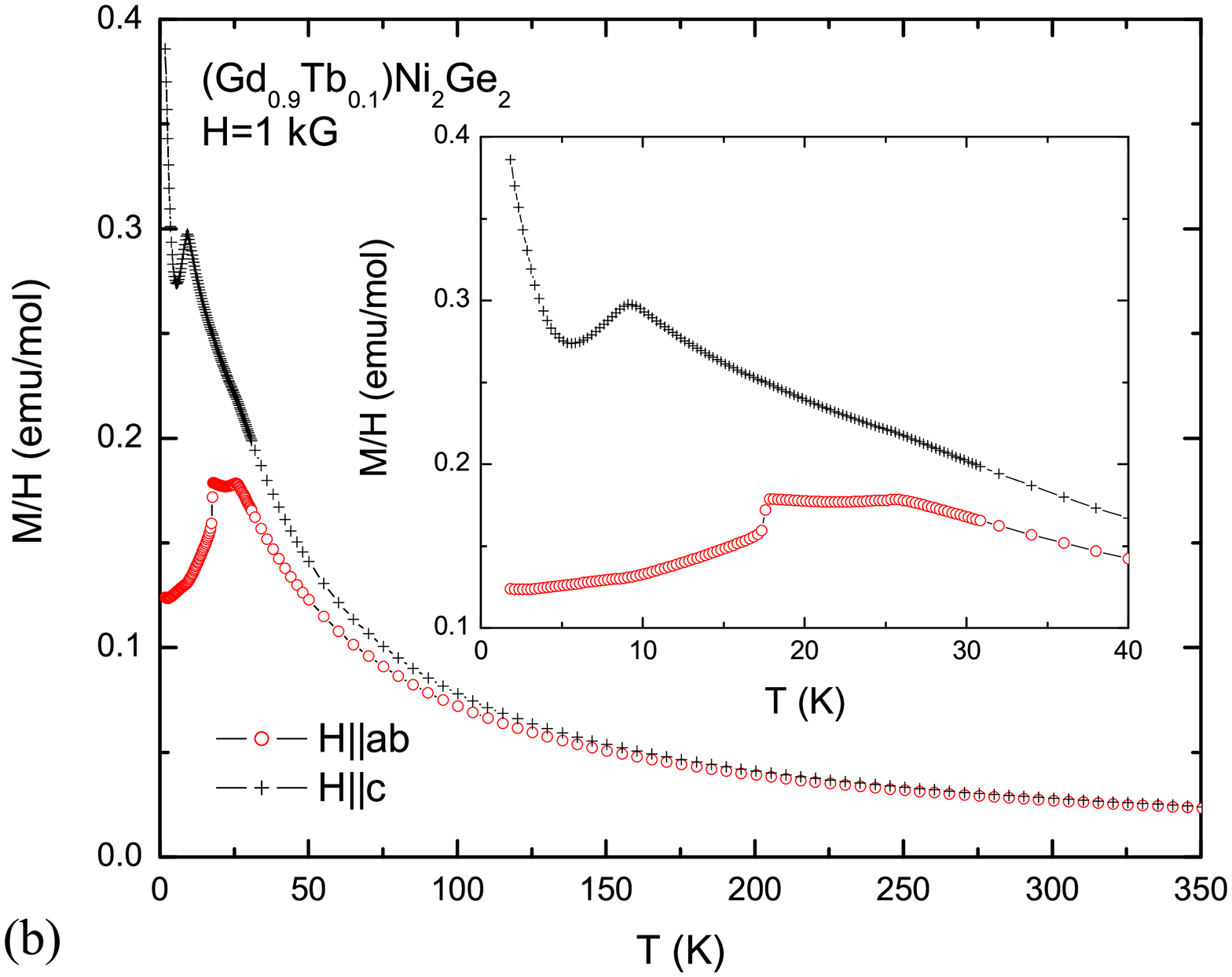}
  \end{minipage}
  \caption{Anisotropic, temperature dependent magnetic susceptibility of (Gd$_{0.15}$Tb$_{0.85})$Ni$_{2}$Ge$_{2}$ (a) and (Gd$_{0.9}$Tb$_{0.1})$Ni$_{2}$Ge$_{2}$ (b).}
\end{figure}

\begin{figure}[htbp]
    \centering
    \includegraphics[width=6.8cm]{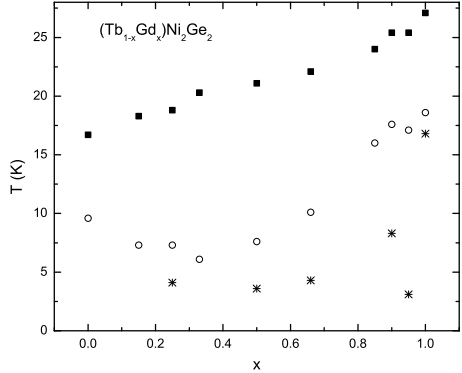}
    \caption{Transition temperature versus x phase diagram for (Tb$_{1-x}$Gd$_{x})$Ni$_{2}$Ge$_{2}$ series.}
\end{figure}

Figure 5 presents the anisotropic low temperature \textit{M(H)} isotherms for the (Tb$_{1-x}$Gd$_{x}$)Ni$_{2}$Ge$_{2}$ series and figure 6 presents the critical fields as a function of \textit{x}.  As can be seen in figures 5a and 6a, when Tb is added to GdNi$_{2}$Ge$_{2}$, the metamagnetic transition seen for field applied in the basal plane drops and disappears for \textit{x} $<$ 0.66.  A somewhat more complex behavior is found for the field applied along the \textit{c}-axis, figures 5b and 6b.  Whereas for \textit{x} $\leq$ 0.5, there are clear transitions near 2 T and 4-5 T, for \textit{x} $>$ 0.85 the upper transition is lost and even the lower transition drops to zero.  Detailed neutron or magnetic X$-$ray diffraction measurements are needed to identify specific regions of figures 4 and 6 with specific ordering wave vectors.

\begin{figure}[htbp]
  \centering
  \begin{minipage}[b]{6.8 cm}
    \includegraphics[width=6.8cm]{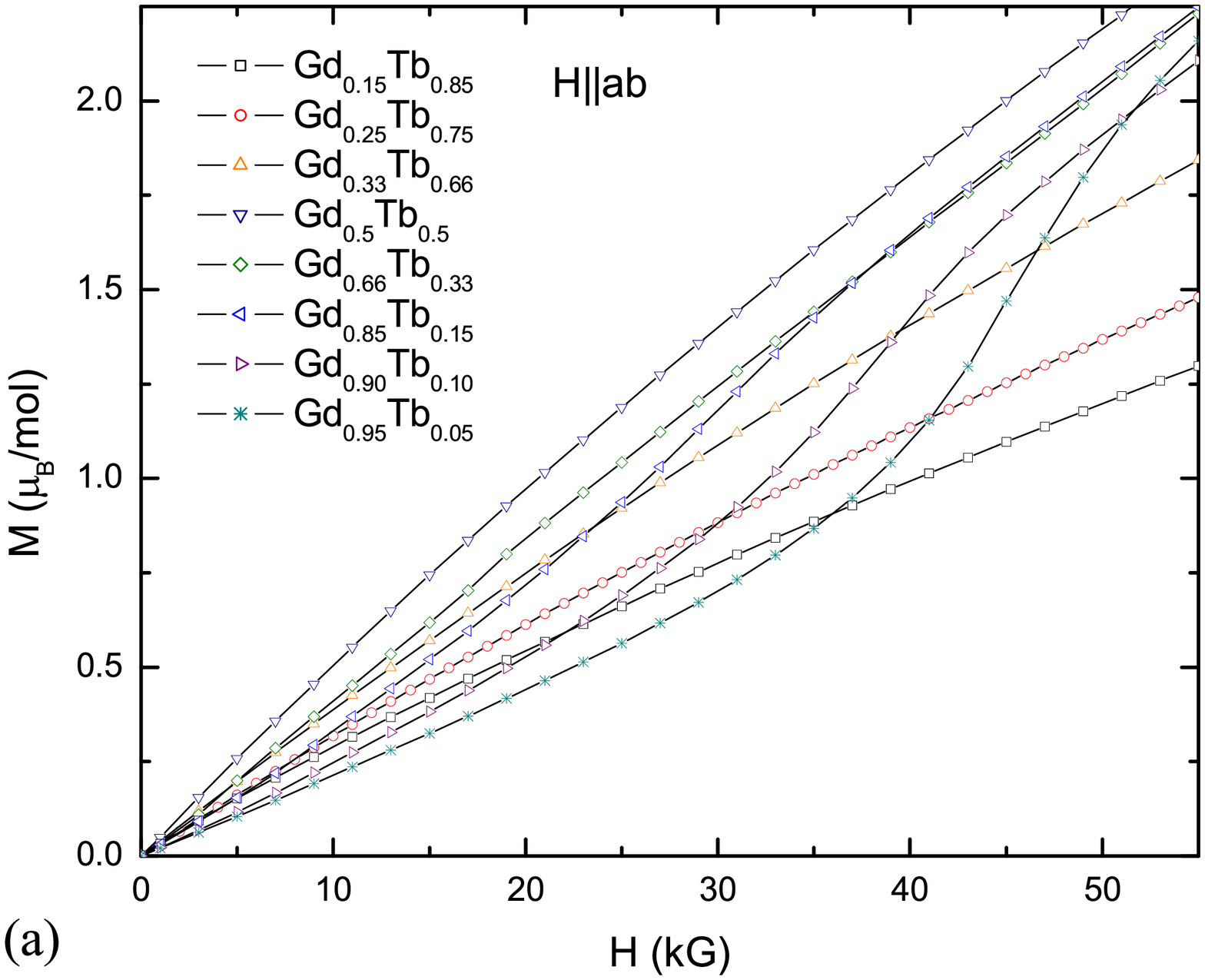}
  \end{minipage}
  \begin{minipage}[b]{6.8 cm}
    \includegraphics[width=6.8cm]{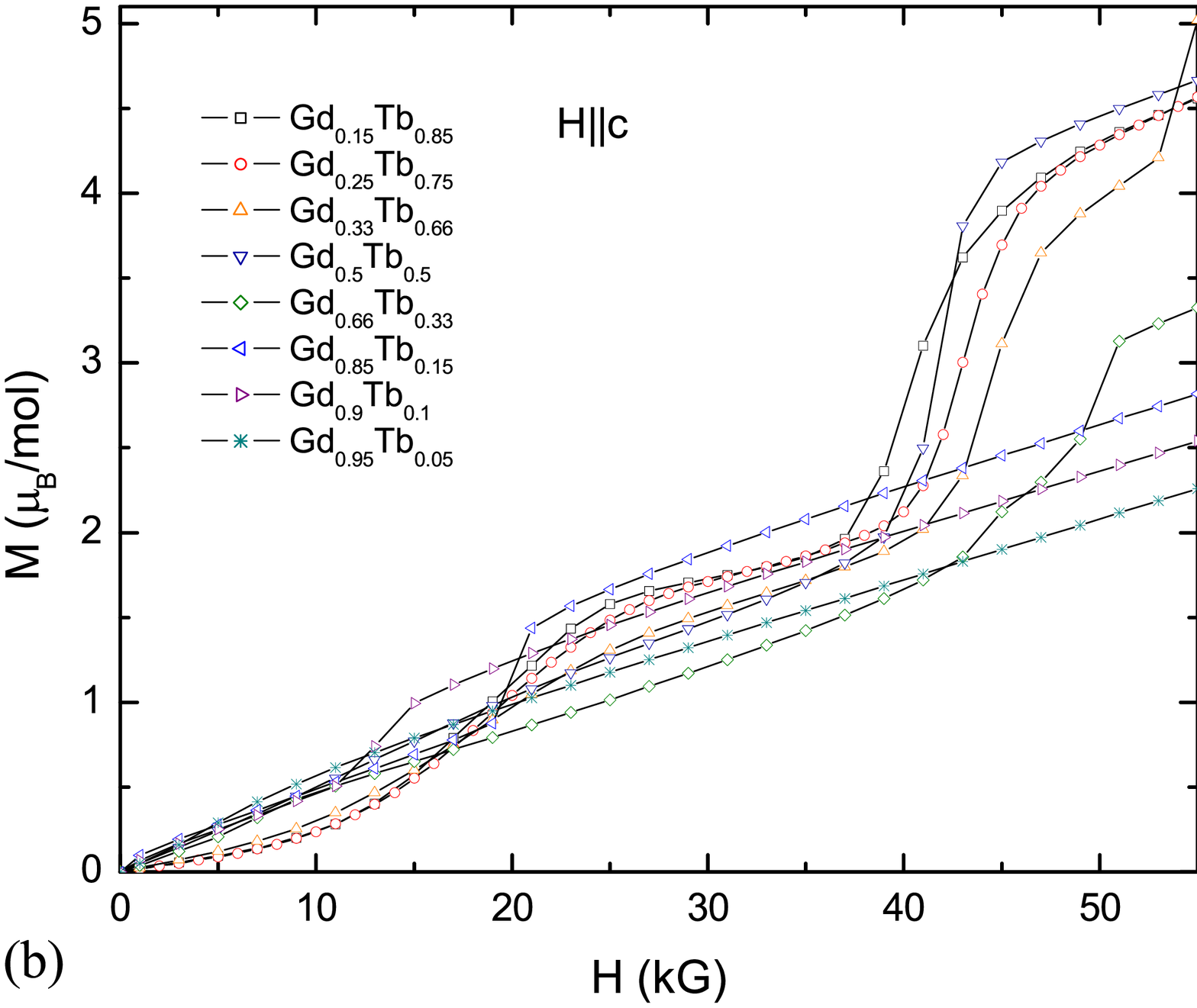}
  \end{minipage}
  \caption{Anisotropic, low temperature, field dependent magnetization of (Tb$_{1-x}$Gd$_{x})$Ni$_{2}$Ge$_{2}$ for field in-plane (a) and field perpendicular to plane (b).}
\end{figure}

\begin{figure}[htbp]
  \centering
  \begin{minipage}[b]{6.8 cm}
    \includegraphics[width=6.8cm]{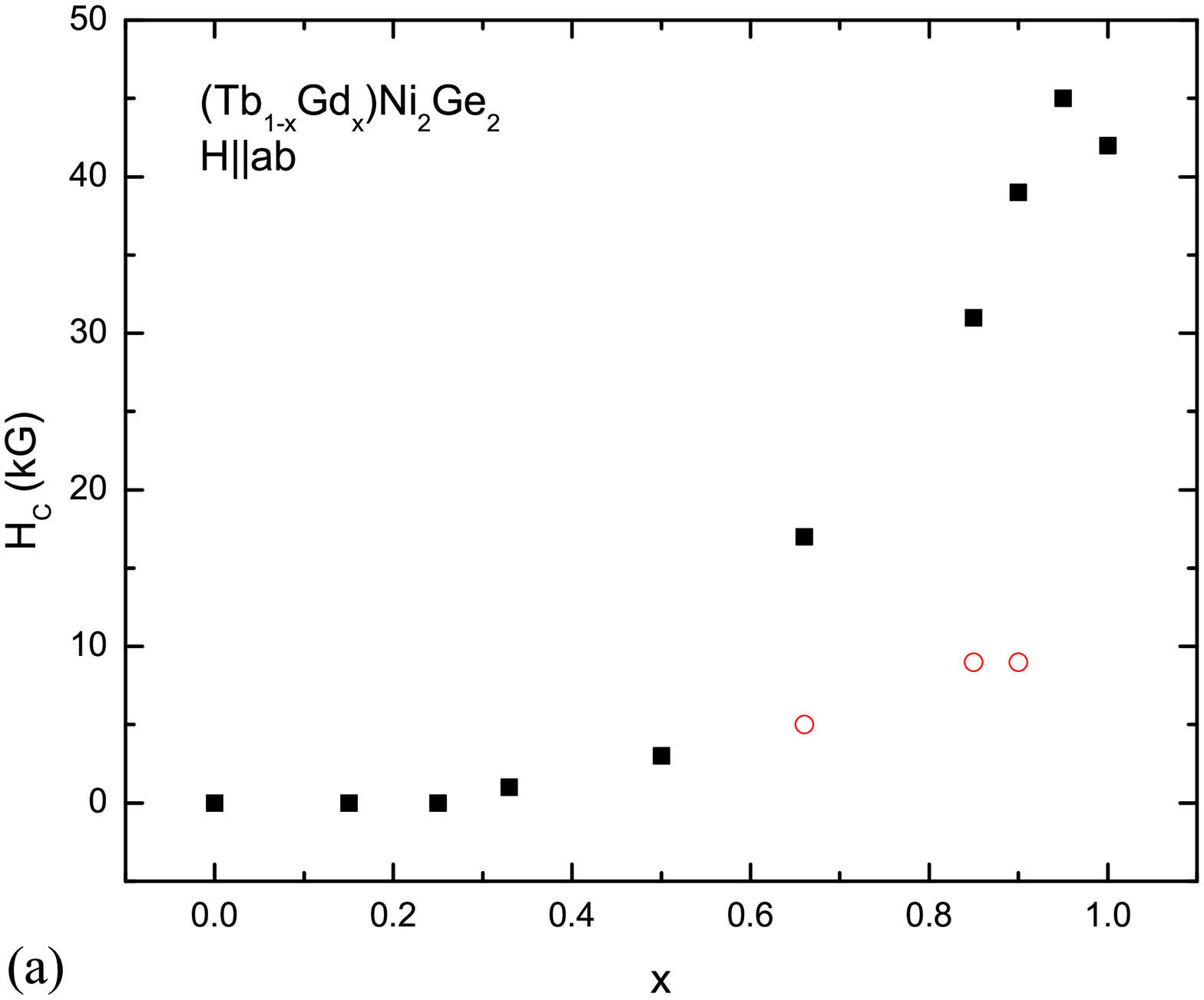}
  \end{minipage}
  \begin{minipage}[b]{6.8 cm}
    \includegraphics[width=6.8cm]{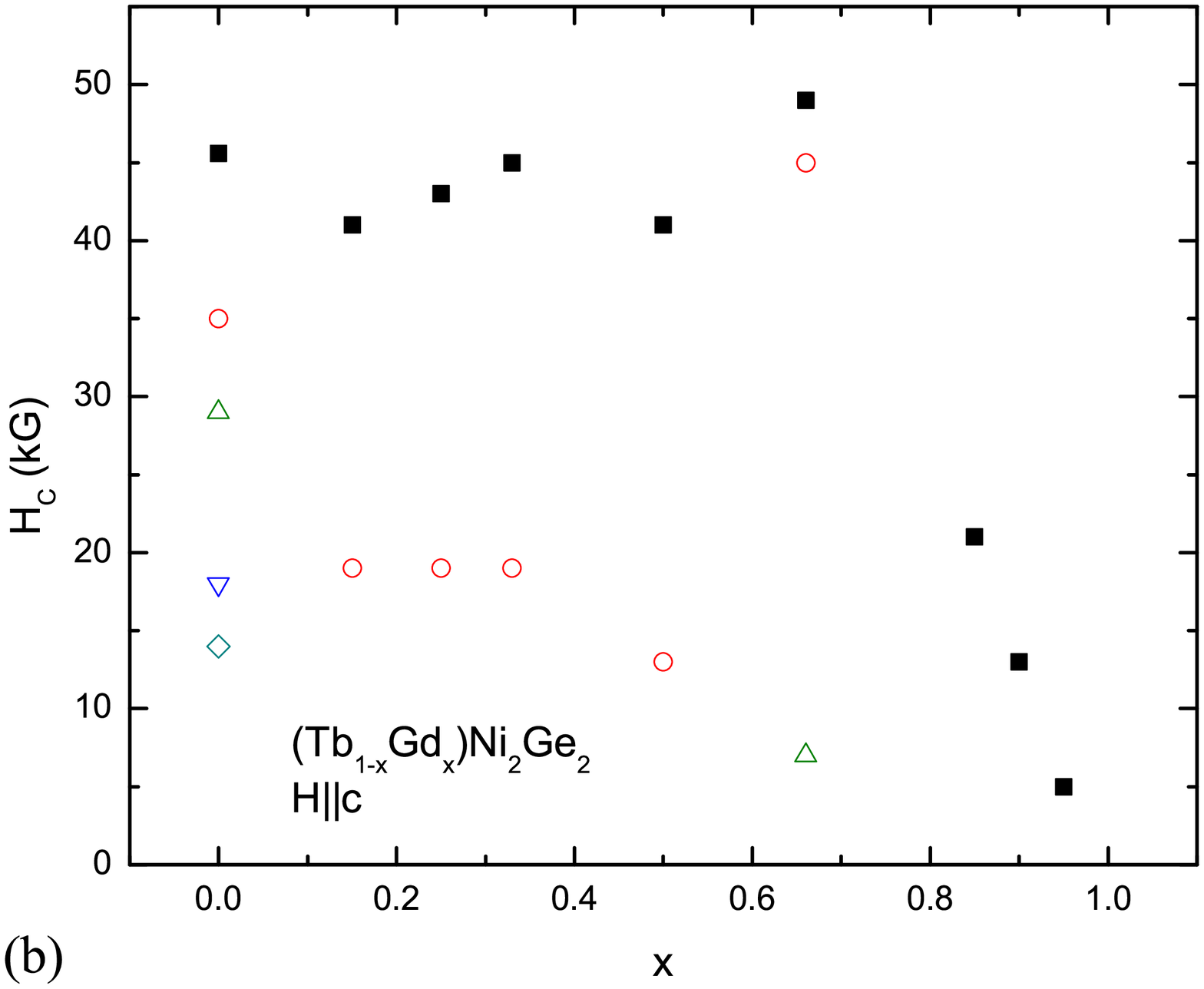}
  \end{minipage}
  \caption{Anisotropic critical field versus x phase diagram for (Tb$_{1-x}$Gd$_{x})$Ni$_{2}$Ge$_{2}$ for field in-plane (a) and field perpendicular to plane (b).}
\end{figure}

Although the ordering temperatures for the (Tb$_{1-x}$Gd$_{x}$)Ni$_{2}$Ge$_{2}$ series scale well with the de Gennes factor (see figure 9 below), this series can only probe de Gennes values between those of pure TbNi$_{2}$Ge$_{2}$ and pure GdNi$_{2}$Ge$_{2}$ (i.e. 10.50 $<$ \textit{dG} $<$ 15.75).  In order to probe a wider range of de Gennes values and to further explore the response of the TbNi$_{2}$Ge$_{2}$ system to a range of perturbations, a related series, (Tb$_{0.75}$R$_{0.25}$)Ni$_{2}$Ge$_{2}$, was studied.  The \textit{M(T)} data for field applied along the \textit{c}-axis for these compounds are shown in figure 7.  Their de Gennes factor ranges from $\sim$7.8 for R = Lu, Y to $\sim$11.8 for R = Gd.  As shown in figure 9 below, de Gennes scaling of \textit{T$_{N}$} continues to hold.  Figure 8 presents the anisotropic, low temperature magnetization for the members of this series.  Regardless of which heavy R is substituted for Tb, the (Tb$_{0.75}$R$_{0.25}$)Ni$_{2}$Ge$_{2}$ samples manifest clear metamagnetism for \textit{H $||$ c} near 4-5 T.  Other transitions are seen near 3 T and/or 1 T.  For \textit{H $||$ ab} there are no clear metamagnetic transitions for \textit{H} $<$ 5.5 T.  The data for R = Tm and Er appear to be clearly offset from the data for R = Y, Lu at high fields for in-plane magnetization.  This raises the question of whether the planar moments associated with Er and Tm are somehow decoupled from the ordering Tb ions.  In an attempt to see if the minority rare earth is ordered or acting as a paramagnetic impurity that is simply undergoing Brillouin saturation, the \textit{H $||$ ab} magnetization was decomposed into a part associated with the Tb sublattice and an additional part, presumably associated with the minority rare earth.  The inset to figure 8b presents the \textit{M(H)} data for (Tb$_{0.75}$Er$_{0.25}$)Ni$_{2}$Ge$_{2}$ and (Tb$_{0.75}$Tm$_{0.25}$)Ni$_{2}$Ge$_{2}$ with the \textit{M(H)} data for (Tb$_{0.75}$Y$_{0.25}$)Ni$_{2}$Ge$_{2}$ subtracted.  Although the data approach saturation for \textit{H}$>$2.5 T, the saturated moment is half that associated with the paramagnetic planar moments. \cite{1}  These results imply that the Er and Tm are indeed taking part in the ordered state, a fact that is supported by magnetic X-ray diffraction studies carried out on other R-mixtures in this family. \cite{11},\cite{12}

\begin{figure}[htbp]
    \centering
    \includegraphics[width=6.8cm]{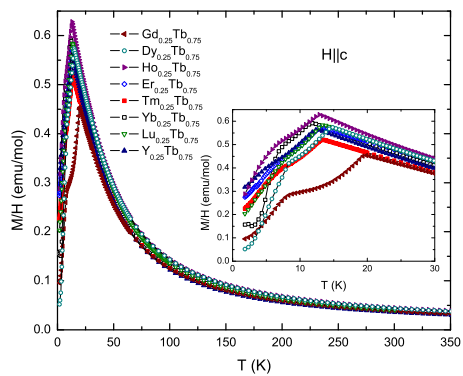}
    \caption{Temperature dependent magnetic susceptibility of (Tb$_{0.75}$R$_{0.25})$Ni$_{2}$Ge$_{2}$ compounds for field applied along the c-axis.}
\end{figure}

\begin{figure}[htbp]
  \centering
  \begin{minipage}[b]{6.8 cm}
    \includegraphics[width=6.8cm]{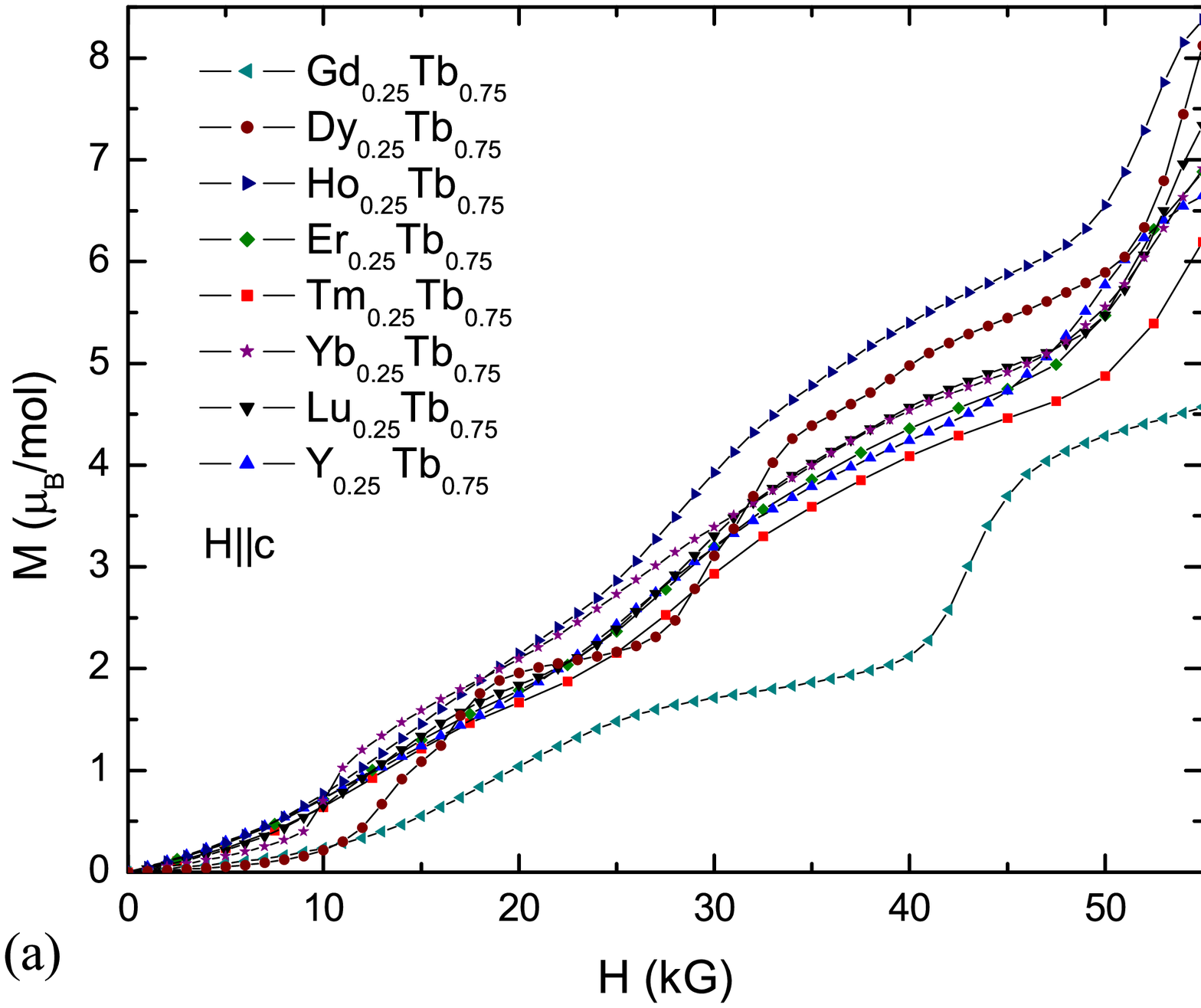}
  \end{minipage}
  \begin{minipage}[b]{6.8 cm}
    \includegraphics[width=6.8cm]{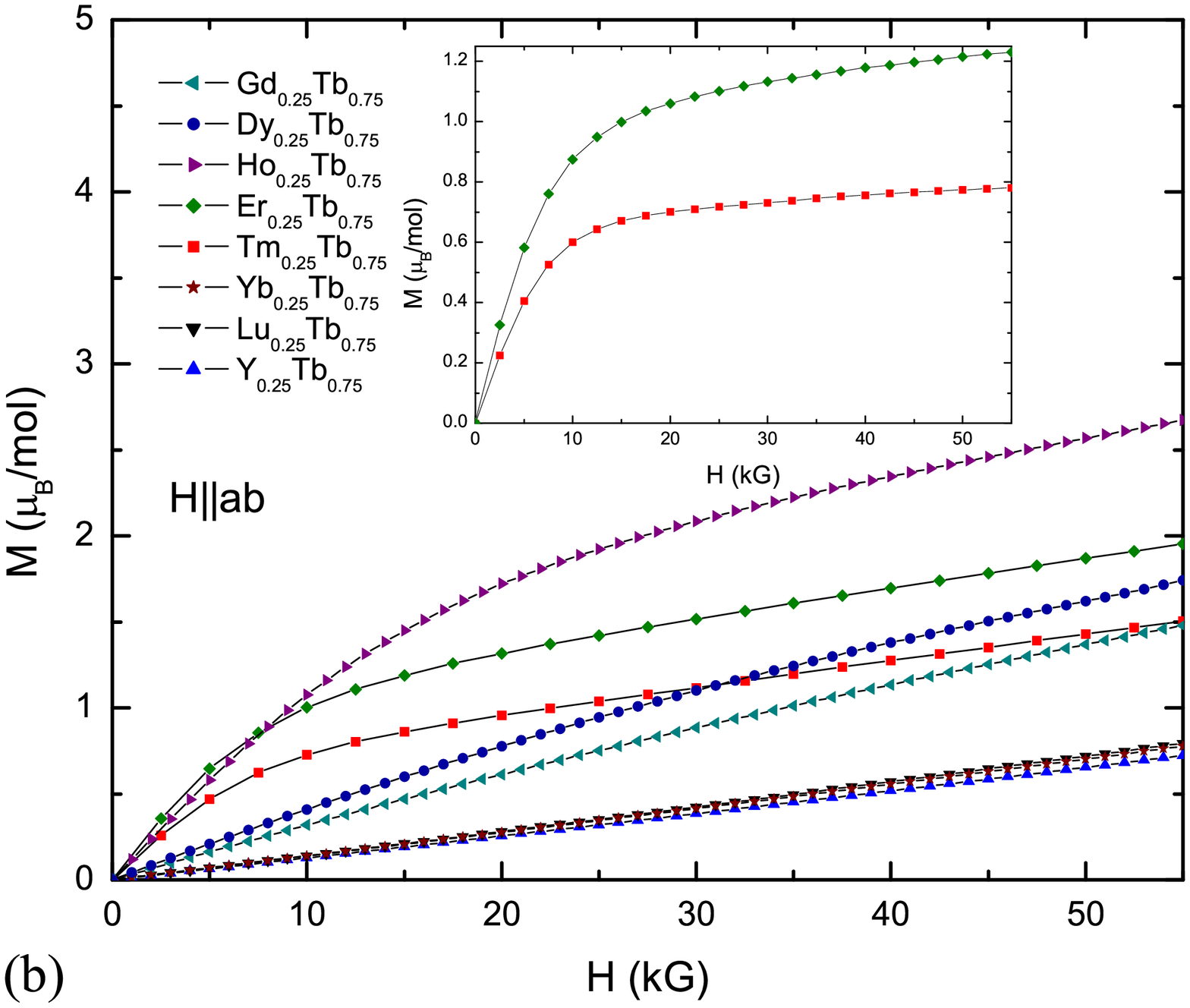}
  \end{minipage}
  \caption{Anisotropic, low temperature magnetization for (Tb$_{0.75}$R$_{0.25})$Ni$_{2}$Ge$_{2}$ compounds for field applied perpendicular to plane (a) and field applied in-plane (b). Inset to figure 8b presents the M(H) data for (Tb$_{0.75}$Er$_{0.25})$Ni$_{2}$Ge$_{2}$ and (Tb$_{0.75}$Tm$_{0.25})$Ni$_{2}$Ge$_{2}$ with the M(H) data for (Tb$_{0.75}$Y$_{0.25})$Ni$_{2}$Ge$_{2}$ subtracted.}
\end{figure}

Lower values of the de Gennes parameter were obtained with the (Gd$_{1-x}$Er$_{x}$)Ni$_{2}$Ge$_{2}$, (R$_{0.5}$Er$_{0.5}$)Ni$_{2}$Ge$_{2}$ and (R$_{0.5}$Tm$_{0.5}$)Ni$_{2}$Ge$_{2}$ series.  The \textit{T$_{N}$} values for these samples, the series discussed above, and those for pure RNi$_{2}$Ge$_{2}$ and (Y$_{1-x}$Tb$_{x}$)Ni$_{2}$Ge$_{2}$ samples from refs. 1 and 5 are presented in figure 9.  The \textit{T$_{N}$} values for three ternary R-mixtures, one quaternary R-mixture, and the m\'elange sample are also shown.   All of these data scale quite well, indicating that sample averaged de Gennes value is a good predictor of \textit{T$_{N}$} for the heavy R members of the (R,R',R''...)Ni$_{2}$Ge$_{2}$ series.

\begin{figure}[htbp]
    \centering
    \includegraphics[width=15cm]{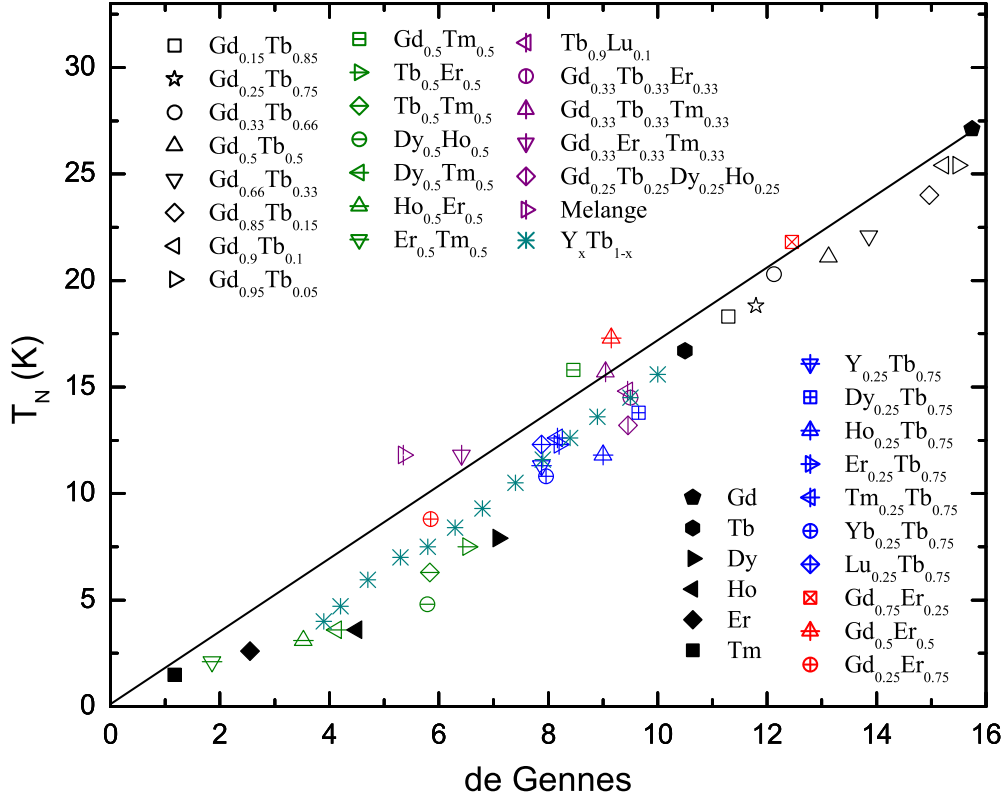}
    \caption{\textit{T$_{N}$} versus average de Gennes factor for (R,R',R''...)Ni$_{2}$Ge$_{2}$ compounds from Table 1, pure heavy rare earth compounds from ref. 1, and (Tb$_{1-x}$Y$_x$)Ni$_2$Ge$_2$ compounds from ref. 5.}
\end{figure}

\section{Discussion}

Although the RNi$_{2}$Ge$_{2}$ series manifests extremes in local moment anisotropy (ranging from axial through
isotropic to planar), the magnetic ordering temperature does not appear to depend too strongly on this anisotropy.
(This makes it different from, for example, the RRh$_{4}$B$_{4}$ series in which extreme axial anisotropy for R =
Tb is thought to be responsible for a higher ordering temperature than that found for R = Gd. \cite{13})  Although
this effect can be seen by observing that the \textit{T$_{N}$} values for the pure, heavy rare earth
RNi$_{2}$Ge$_{2}$ compounds do not significantly deviate from de Gennes scaling (i.e. \textit{T$_{N}$} for the
axial R = Tb is not higher than that for the isotropic R = Gd), this relative insensitivity of \textit{T$_{N}$} to
local moment anisotropy can be further illustrated by examining mixtures which have similar average de Gennes
values and dramatically different net anisotropies.  A particularly clear example of this can be found by
examining data for the (Tb$_{0.9}$Lu$_{0.1}$)Ni$_{2}$Ge$_{2}$,
(Gd$_{0.25}$Tb$_{0.25}$Dy$_{0.25}$Ho$_{0.25}$)Ni$_{2}$Ge$_{2}$, and (Gd$_{0.5}$Tm$_{0.5}$)Ni$_{2}$Ge$_{2}$
samples.  These samples are part of the cluster of data points in figure 9 centered near \textit{dG} $\sim$ 9 and
\textit{T$_{N}$} $\sim$ 15 K.  Figure 10 presents the anisotropic, low temperature magnetic susceptibility data
for each of these samples.  (Tb$_{0.9}$Lu$_{0.1}$)Ni$_{2}$Ge$_{2}$ is extremely axial (which is expected, given
that it is a slight non-magnetic dilution away from the pure, Ising-like TbNi$_{2}$Ge$_{2}$),
(Gd$_{0.25}$Tb$_{0.25}$Dy$_{0.25}$Ho$_{0.25}$)Ni$_{2}$Ge$_{2}$ is only weakly axial (actually approaching isotropy in the paramagnetic state),
and (Gd$_{0.5}$Tm$_{0.5}$)Ni$_{2}$Ge$_{2}$ is manifesting the planar anisotropy associated with pure
TmNi$_{2}$Ge$_{2}$.  Despite these clear and large differences in local moment anisotropy, the de Gennes factor
appears to be the primary agent in determining the antiferromagnetic ordering temperature.

\begin{figure}[htbp]
  \centering
  \begin{minipage}[b]{6.8 cm}
    \includegraphics[width=6.8cm]{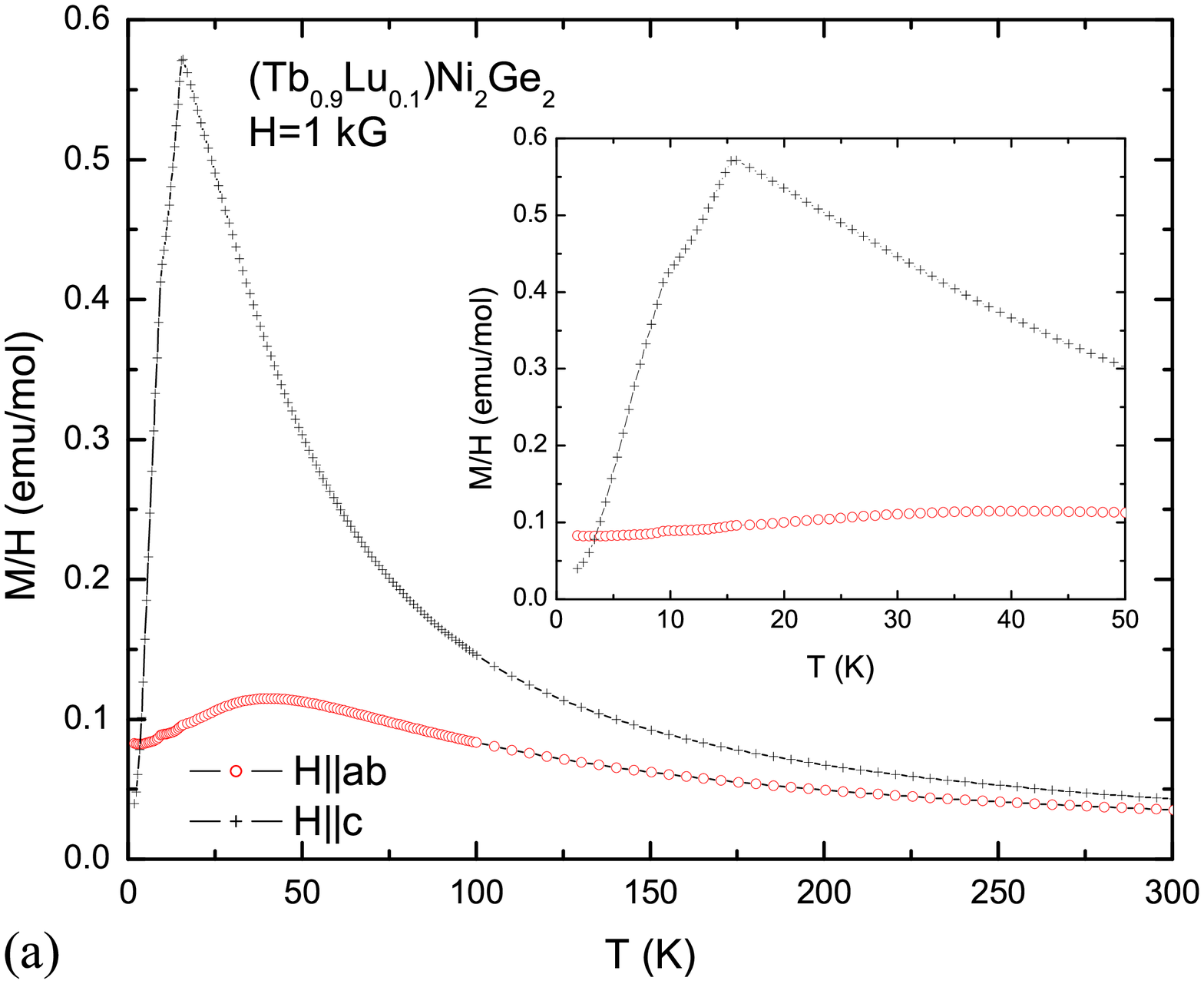}
  \end{minipage}
  \begin{minipage}[b]{6.8 cm}
    \includegraphics[width=6.8cm]{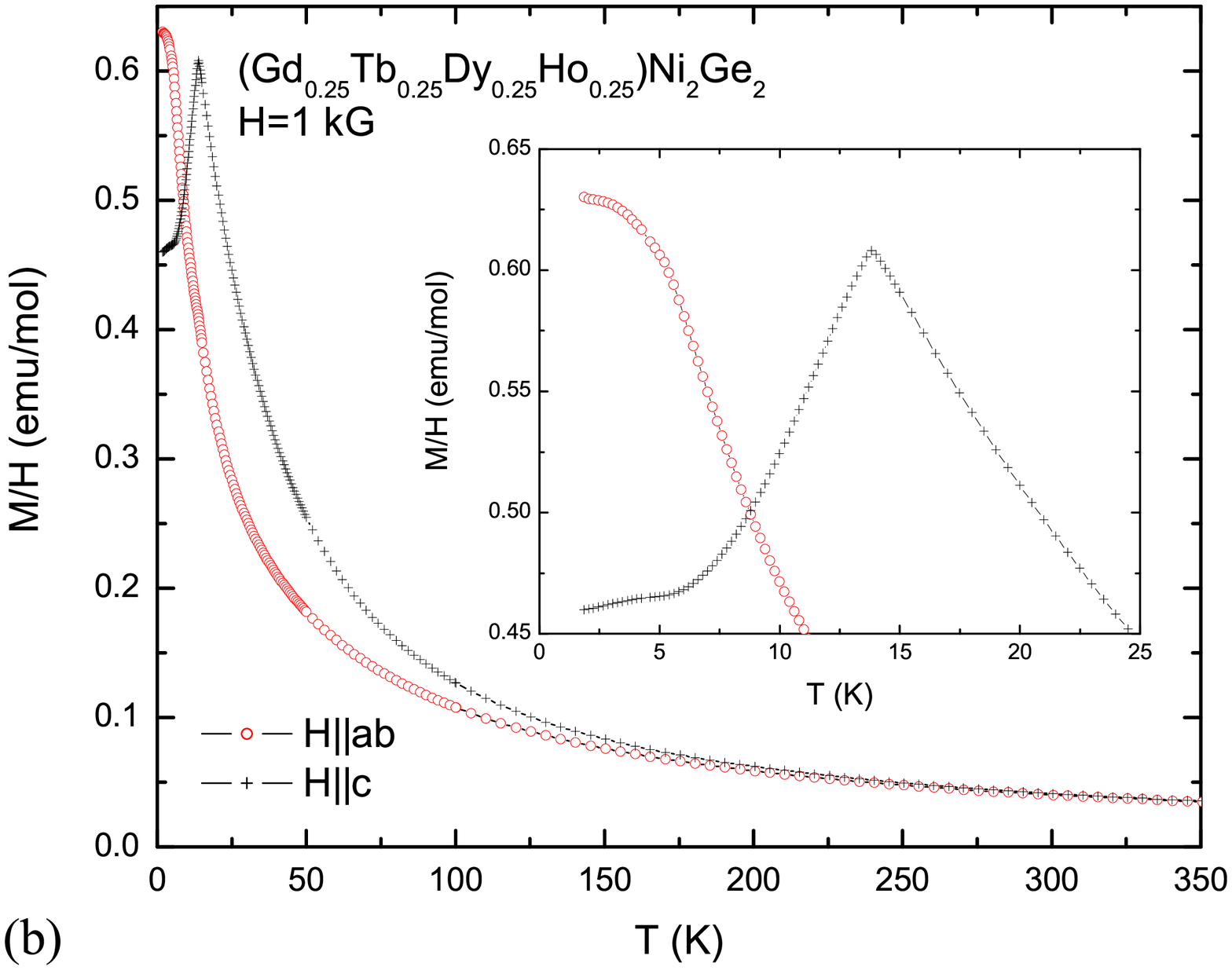}
  \end{minipage}
  \begin{minipage}[b]{6.8 cm}
    \includegraphics[width=6.8cm]{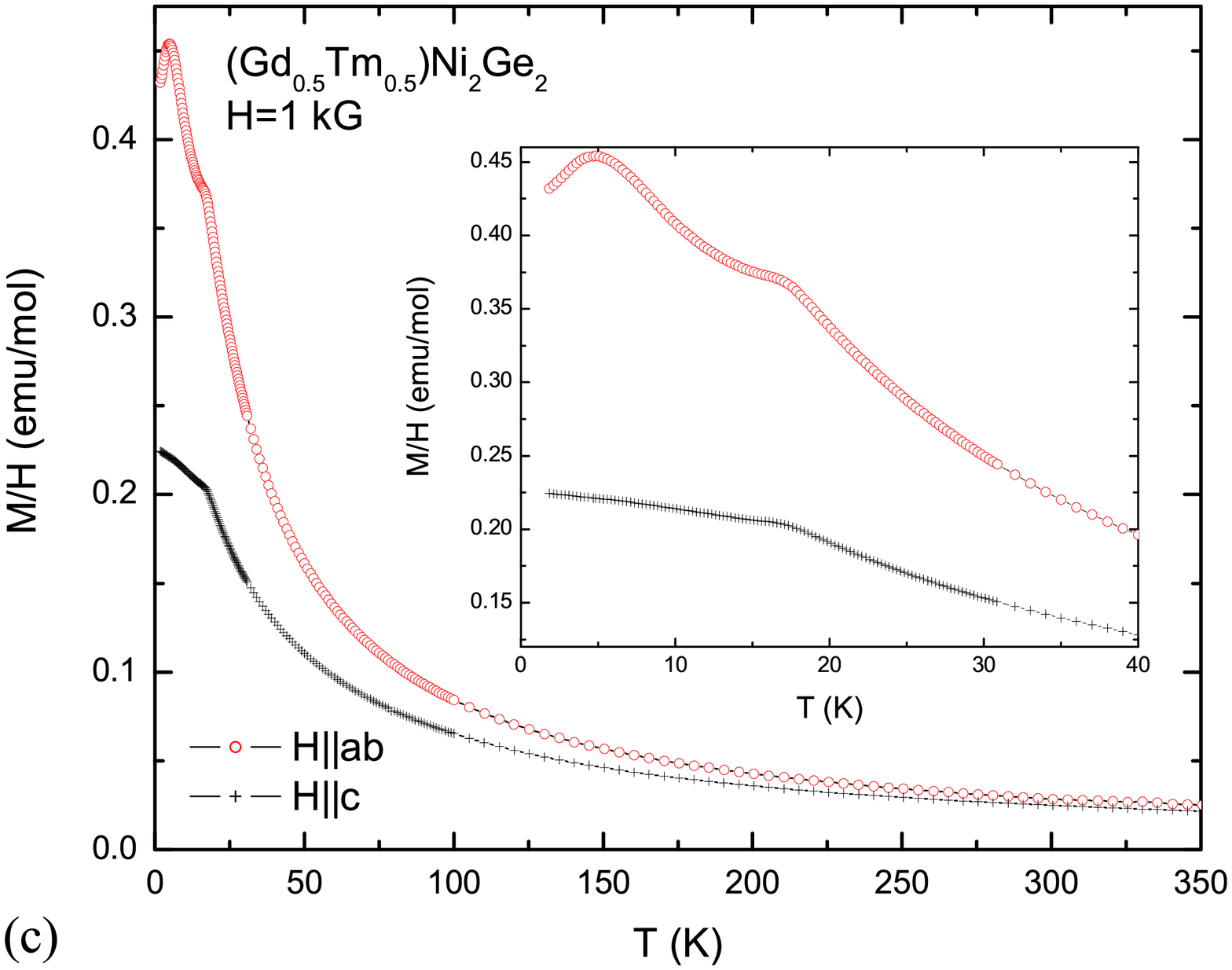}
  \end{minipage}
  \caption{Anisotropic low temperature magnetic susceptibility for the (Tb$_{0.9}$Lu$_{0.1}$)Ni$_{2}$Ge$_{2}$ (a), (Gd$_{0.25}$Tb$_{0.25}$Dy$_{0.25}$Ho$_{0.25}$)Ni$_{2}$Ge$_{2}$(b), and (Gd$_{0.5}$Tm$_{0.5}$)Ni$_{2}$Ge$_{2}$ (c) samples.}
\end{figure}

One feature that contributes to the RNi$_{2}$Ge$_{2}$ series' compliance with de Gennes scaling is that so many of the heavy R members order with a similar type of wave vector:  (0, 0, \textit{l}) with \textit{l}  ranging from 0.75 to 0.81.  This trend continues for the samples containing multiple rare earths with (Gd$_{0.25}$Tb$_{0.25}$Dy$_{0.25}$Ho$_{0.25}$)Ni$_{2}$Ge$_{2}$ and (Gd$_{0.33}$Er$_{0.33}$Tm$_{0.33}$)Ni$_{2}$Ge$_{2}$ having \textit{l} = 0.752 and 0.759, respectively, \cite{12} and the m\'elange sample with its nine moment-bearing rare earths (including several light ones) having \textit{l} = 0.755 \cite{11}.

\section{Acknowledgements}

We would like to acknowledge Nate Kelso for his seminal contributions at the inception of this work. We would also
like to thank H. S. Thompson and H. Humbert for their moral support. Ames Laboratory is operated for the U.S. Department of Energy by Iowa State University under Contract No. W-7405-Eng.-82. This work was supported by the Director for Energy Research, Office of Basic Energy Sciences.

\label{}



\end{document}